\documentclass[9pt]{article}  \usepackage{times}
\usepackage{graphicx, textgreek}

\usepackage{color}

\usepackage[round,numbers,sort&compress]{natbib} 
\usepackage{amsmath}
\usepackage{nccmath}   
\usepackage{amssymb}
\usepackage{stackrel}
\usepackage{chemarrow}
\usepackage{graphicx}
\usepackage{textgreek}

\usepackage{booktabs}
\usepackage{dcolumn}
\usepackage{rotating}
\usepackage{multirow}

\begin{document}

	\twocolumn[{\LARGE \textbf{The excitable fluid mosaic\\*[0.2cm]}}
	{\large Thomas Heimburg$^\ast$\\*[0.1cm]
		{\small Niels Bohr Institute, University of Copenhagen, Blegdamsvej 17, 2100 Copenhagen \O, Denmark}\\*[-0.1cm]

		{\normalsize \textbf{ABSTRACT}\hspace{0.5cm} The Fluid Mosaic Model by Singer \& Nicolson proposes that biological membranes consist of a fluid lipid layer into which integral proteins are embedded. The lipid membrane acts as a two-dimensional liquid in which the proteins can diffuse and interact. Until today, this view seems very reasonable and is the predominant picture in the literature. However, there exist broad melting transitions in biomembranes some 10--20 degrees below physiological temperatures that reach up to body temperature. Since they are found below body temperature, Singer \& Nicolson did not pay any further attention to the melting process. But this is a valid view only as long as nothing happens. The transition temperature can be influenced by membrane tension, pH, ionic strength and other variables. Therefore, it is not generally correct that the physiological temperature is above this transition. The control over the membrane state by changing the intensive variables renders the membrane as a whole excitable. One expects phase behavior and domain formation that leads to protein sorting and changes in membrane function. Thus, the lipids become an active ingredient of the biological membrane. The melting transition affects the elastic constants of the membrane. This allows for the generation of propagating pulses in nerves and the formation of ion-channel-like pores in the lipid membranes. Here we show that on top of the fluid mosaic concept there exists a wealth of excitable phenomena that go beyond the original picture of Singer \& Nicolson.		\\*[0.3cm] }}
	\noindent\footnotesize{\textbf{Keywords:} thermodynamics; domains; rafts ; elastic constants; ion channels; nerves\\*[0.1cm]}
	\noindent\footnotesize {$^{\ast}$corresponding author, theimbu@nbi.ku.dk. }\\
	\vspace{0.3cm}
	]

	\normalsize


\section{Introduction}
\label{introduction}

The ``Fluid Mosaic Model'' by Singer and Nicolson \cite{Singer1972} is a landmark publication that has shaped the view of the role and function of the biological membrane until today. It assumes that the biological membrane is a fluid lipid layer containing integral proteins. Peripheral (or soluble) proteins are not considered in much detail, because their interactions with membranes are considered being weak. The authors call it a thermodynamic model because it is based on minimizing the free energy of hydrophilic and hydrophobic contacts. The hydrophobic effect was pioneered by Walter Kauzmann in the 1950s \cite{Kauzmann1959}. It explores the interactions of hydrophobic groups, e.g., the lipid chains and the hydrophobic amino acids, with water. Kauzmann found that the free energy of the transfer of apolar molecules such as alkanes from an oil-like environment to water is temperature dependent. This implies that the interface of apolar groups with water possesses both an enthalpy and entropy contribution with an associated heat capacity of the hydrophobic interface. The consequence is that apolar groups prefer apolar environments. The free energy of electrostatic interactions is described by the Debye-H\"uckel theory \cite{Debye1923}.

\begin{figure}[htbp]
\centering
\includegraphics[width=250pt]{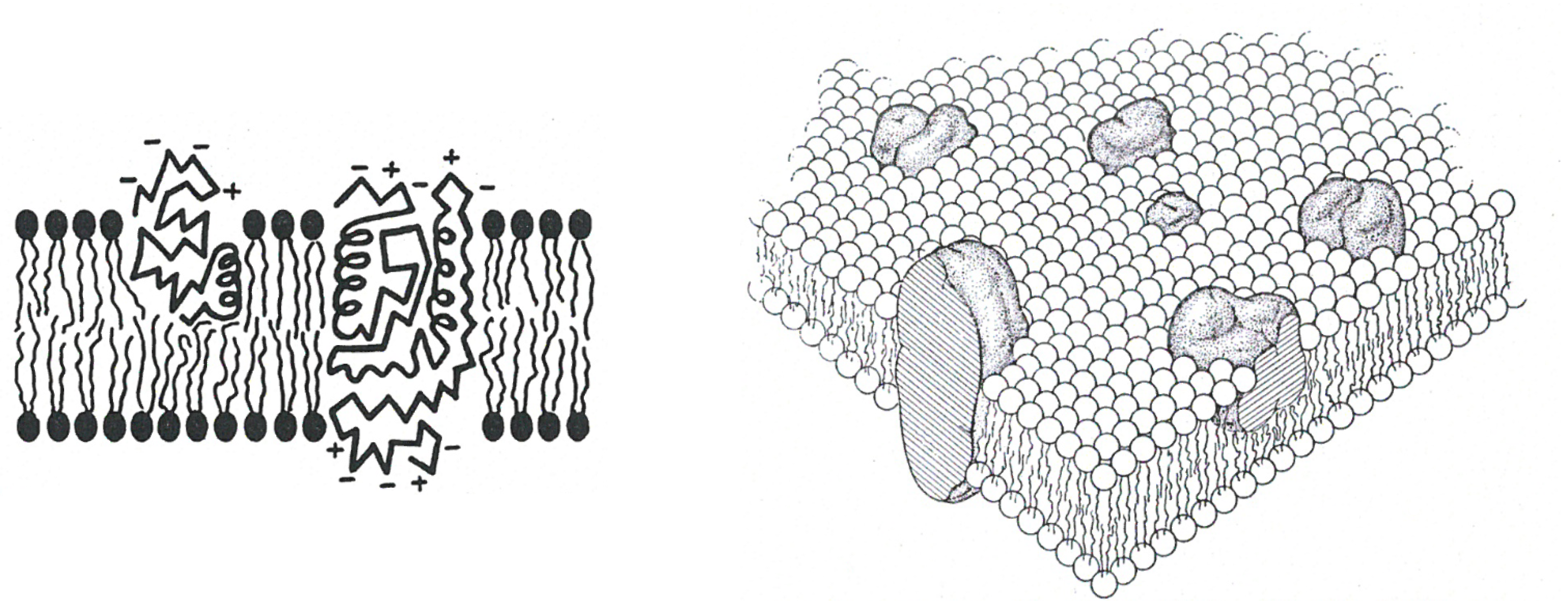}
\caption{The Singer \& Nicolson model \cite{Singer1972} suggests that proteins with hydrophobic parts are embedded into the membrane. The lipid matrix serves a fluid two-dimensional confinement for the integral membrane proteins that may interact in order to generate biological function.}
\label{singernicolson}
\end{figure}

The primary sequence of many proteins contains stretches of amino acids with apolar residues. These residues can avoid contact with water when they are embedded into the hydrophobic core of the membrane. At the same time, charged amino acids are located on the outside of the membrane in order to optimize the interaction with the electrolyte (see Fig. \ref{singernicolson}, left). When the ``fluid mosaic model'' was published, crystal structures of membrane proteins did not exist. Therefore, the prediction that integral proteins exist and fold in a manner that allows them to minimize the free energy of charges and hydrophobic residues at the same time was very important.

\begin{figure}[htbp]
\centering
\includegraphics[width=250pt]{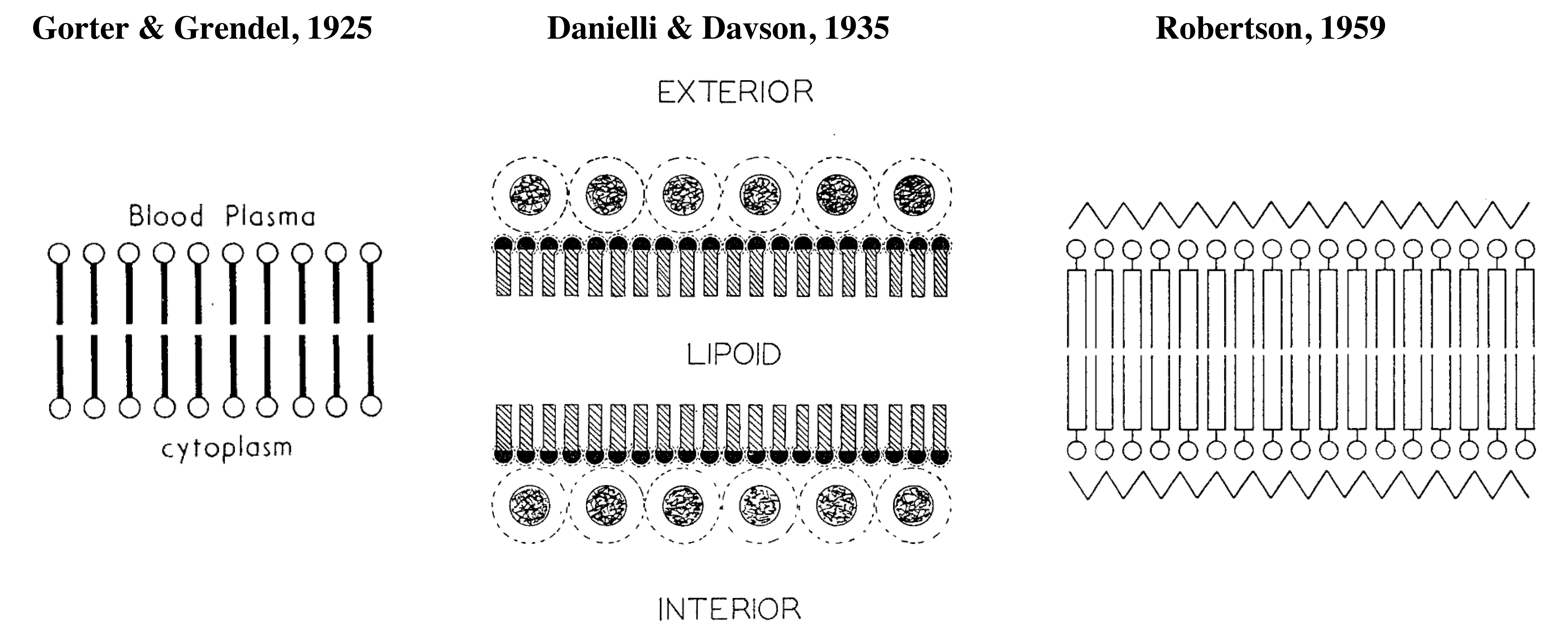}
\caption{Earlier membrane models. Left: Gorter and Grendel first proposed that lipid membranes consist of two monolayers of lipids \cite{Gorter1925} (image from \cite{Robertson1959}. Center: The Danielli \& Davson model of a biological membrane describes the membrane as a oil-like phase with lipids at the interface \cite{Danielli1935}. Globular proteins are adsorbed to the surface. Right: Robertson used electron microscopy to identify the thickness of the biological membrane and proposed that the biological membranes consist of bilayers with a surface layer of proteins \cite{Robertson1959}.}
\label{gortergrendeldaniellidavsonrobertson}
\end{figure}

In 1925, Gorter and Grendel extracted lipids from erythrocytes and deposited them as monolayers on an aqueous surface \cite{Gorter1925}. From the overall lipid area they found, they concluded that the lipid membrane consists of two monolayers of lipids (see Fig. \ref{gortergrendeldaniellidavsonrobertson}, left). They stated: ``It is clear that all our results fit in well with the supposition that the chromocytes are covered by a layer of fatty substances that is two molecules thick.'' This, however, was a statement about the lipid membrane rather that about the biological membrane with all its ingredients. Ten years later, Danielli \& Davson \cite{Danielli1935} proposed that the biological membrane contains oil-like substances (lipoids), lipids and globular proteins. It considered the membrane as a layer of unknown thickness thinner than the microscopic resolution of $\approx$200 nm with proteins adsorbed to its surface (Fig. \ref{gortergrendeldaniellidavsonrobertson}, center). Using the newly developed electron microscopy, Robertson deduced that cell membranes are around 75 \AA, including a protein layer on the surface \cite{Robertson1959}, see Fig. \ref{gortergrendeldaniellidavsonrobertson} (right).

So, we see here what is new in the ``fluid mosaic model''. Rather then considering the membrane as a rigid structure, it suggests that the lipid membrane is a two-dimensional fluid reaction platform for integral proteins.

Singer \& Nicolson noted that biomembranes display melting transitions of their lipids. However, they ruled out that melting is important due to following observation:

\begin{itemize}
\setlength\itemsep{-0.1cm}
\item The biological membrane is found in a fluid state.
\item The lipid composition in the membrane adapts to different growth temperatures. Special enzymes help the membrane to change the lipid composition in order to remain fluid.
\item The fluidity of the membrane is required to maintain translational mobility of the integral proteins.
\item Some lipids (but rather a small fraction) may interact specifically with the integral proteins but generally proteins and lipids diffuse independently and do not interact. Quotes from \cite{Singer1972}:

\begin{quote}
``{\ldots} calorimetric data {\ldots} give no significant indication that the association of proteins with the phospholipids of intact membranes affects the phase transitions of the lipids themselves.''
\end{quote}

\begin{quote}
``Such results therefore suggest that the phospholipids and proteins of membranes do not interact strongly; in fact, they appear to be largely independent''.
\end{quote}

With this statement the authors basically rule out that the lipid-protein interactions or phase behavior of the membrane can affect the distribution of proteins, or mediate protein-protein interactions.

\item  The authors discuss the width of the lipid phase transitions. They mention that the transition is broad and is probably inconsistent with a cooperative unit size of more than 100 lipids. With this statement they foresee the possibility of nanoscopic domains in the biomembrane reminiscent of what later was called ``rafts''. On similar grounds Singer \& Nicolson conclude that there probably is no long-range order within the membrane. Short-range order is thought to be mediated by proteins.

\end{itemize}

The main conclusion of Singer and Nicolson is that (Quote):

\begin{quote}
``The fluid mosaic structure is therefore formally analogous to a two-dimensional oriented solution of integral proteins {\ldots} in the viscous phospholipid bilayer environment''
\end{quote}


\section{Beyond the fluid mosaic model}
\label{beyondthefluidmosaicmodel}


\subsection{Melting of biomembranes, and how it depends on the thermodynamics variables}
\label{meltingofbiomembranesandhowitdependsonthethermodynamicsvariables}

Assuming that something like ``physiological conditions'' (well-defined temperature, pressure, tension, pH, electrostatic potential, {\ldots}) in fact exists, the conclusions of Singer \& Nicolson made in 1972 all seem reasonable and are mostly in agreement with experimental facts. However, it is not generally true that physiological conditions are static. There have been numerous observations that suggest that one should go beyond the fluid mosaic model without stating that it is wrong.

\begin{figure}[htbp]
\centering
\includegraphics[width=200pt]{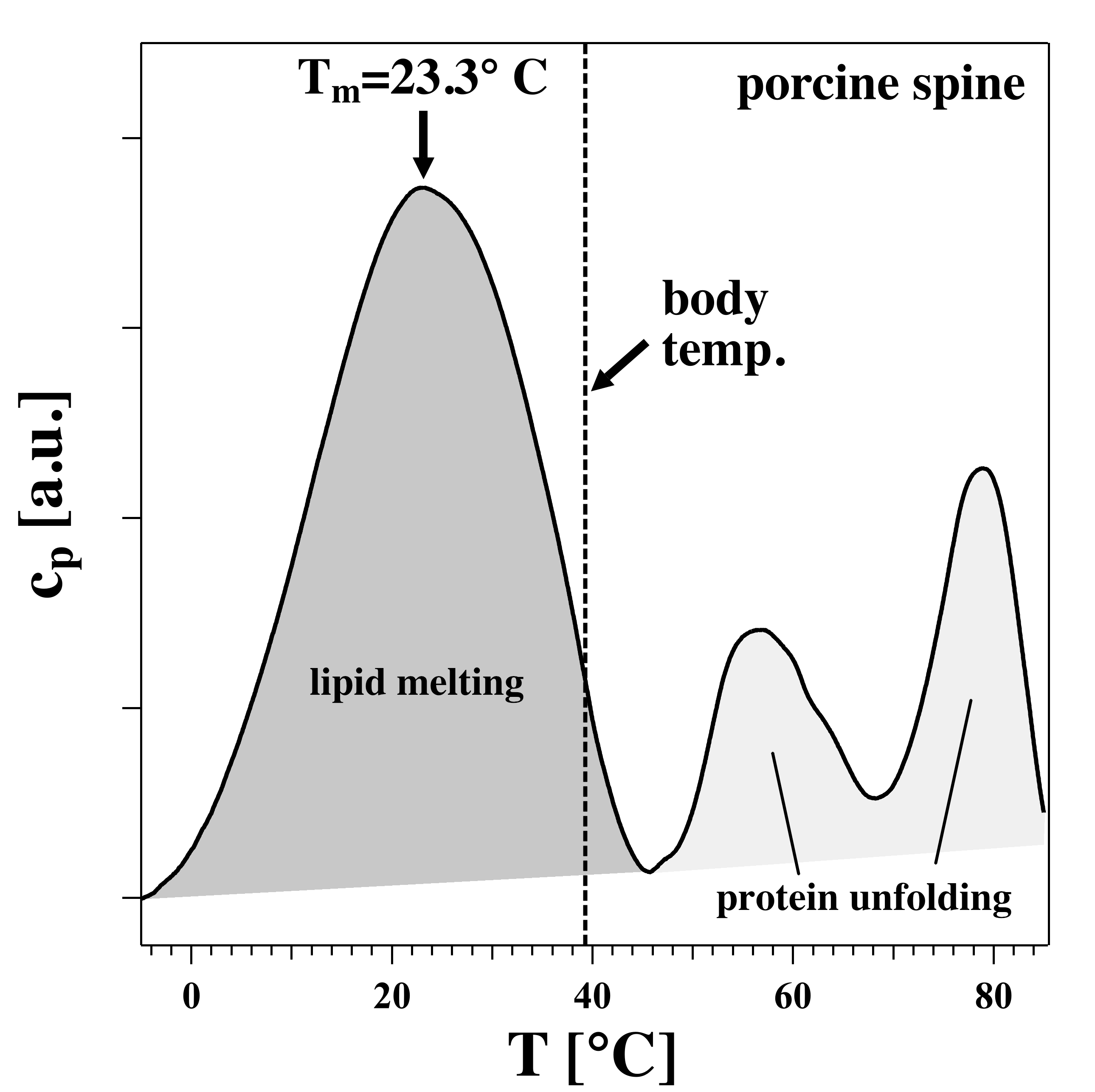}
\caption{Heat capacity profile of lipid melting in membranes from porcine spine. One can clearly recognize the lipid melting peak below body temperature and the protein unfolding peaks above body temperature. Adapted from \cite{Muzic2019}.}
\label{pig_nerve}
\end{figure}

Biological membranes can melt, and they typically do this at temperature about 10--20$^\circ$ below body temperature (see Fig. \ref{pig_nerve}, \cite{Muzic2019, Faerber2022, Fedosejevs2022}. Older literature is reviewed in \cite{Heimburg2007a}). The underlying theme of this review is that membranes are fluid only when nothing happens.

\begin{figure}[htbp]
\centering
\includegraphics[width=200pt]{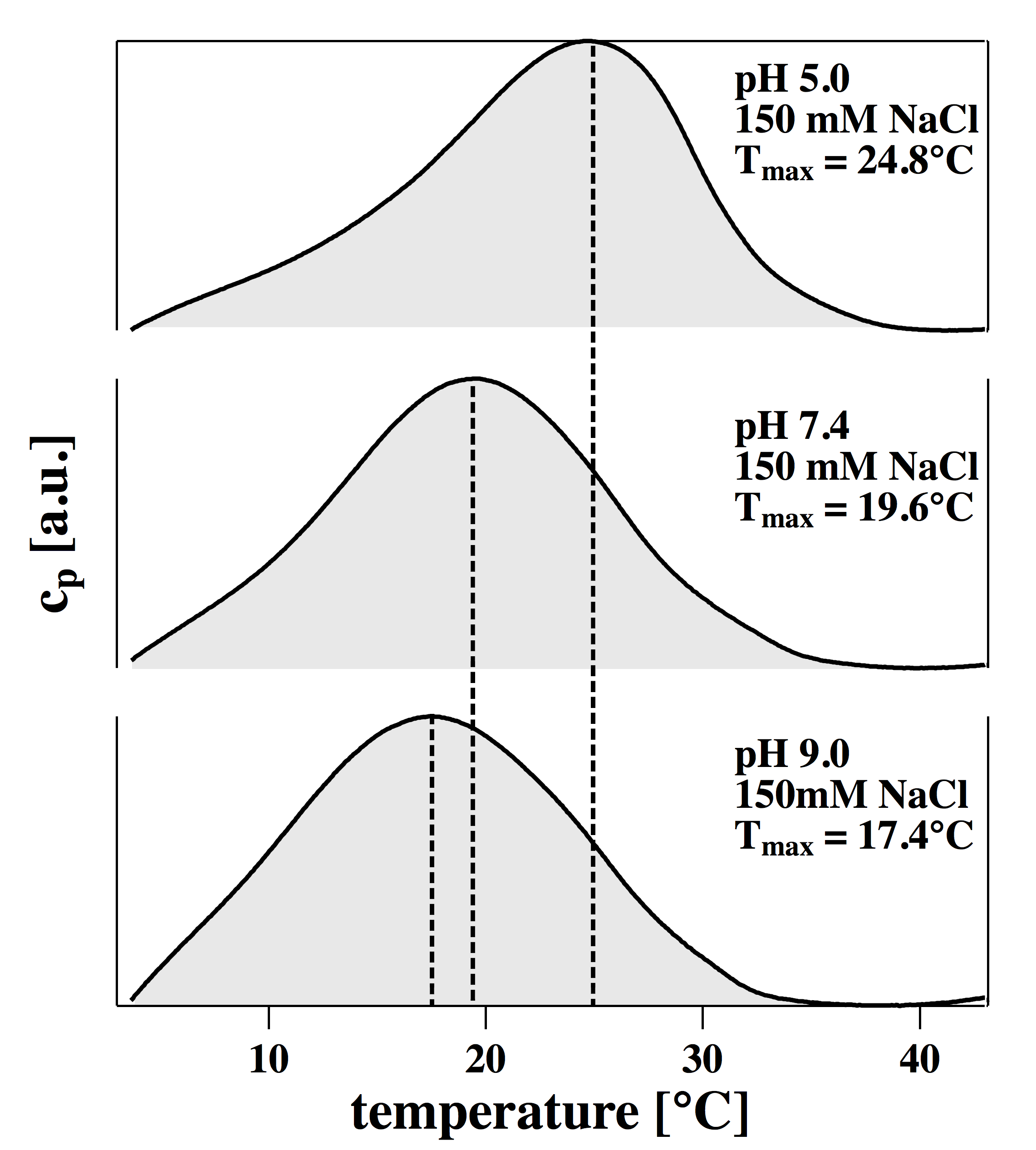}
\caption{Heat capacity profiles of E.coli membranes at different pH values. The transition shifts by 7.4 degrees upon a change in pH from 9 to 5. Adapted from \cite{Muzic2019}.}
\label{muzic2019}
\end{figure}

There exist numerous variations of the environmental conditions of the environment of biological membranes that can change the state of the membrane. We list a few:

\begin{itemize}
\setlength\itemsep{-0.1cm}
\item lowering of temperature will render the membrane solid (see Fig. \ref{pig_nerve}). This can happen after a sudden change in the environmental conditions for cold-blooded animals.
\item pH-changes can increase or lower transition temperatures, both in model membranes and biomembranes \cite{Trauble1976, Muzic2019, Fillafer2021}. A pH change from 9 to 5 causes a shift of the transition maximum by 7.4 degrees upwards (see Fig. \ref{muzic2019}), i.e., $\approx 1.9^\circ$C\slash pH-unit. Larger values of $\approx 5-35^\circ$C\slash pH-unit were discussed in \cite{Fillafer2021}. In a crowded cytosol where diffusion is inhibited \cite{Rivas2016}, pH changes may be local. They may occur, for instance, as a consequence of the activity of ATPases or other hydrolytic enzymes.
\item changes in lateral tension of the membrane also influences transition temperatures and affect the excitability of nerves \cite{Heimburg2022a}. This is important during bending of joints during the movement of animals. Estimates using the properties of artificial lipids would lead to a decrease in transition temperature by 1$^\circ$ when the length of a cylindrical membrane is increased by 4.8\% - corresponding to a lateral tension of -3.3mN\slash m in the lipid layer \cite{Heimburg2022a}.
\end{itemize}

The following variables can also change the state of the membrane:
\begin{itemize}
\setlength\itemsep{-0.1cm}
\item changes in hydrostatic pressure \cite{Ebel2001, Muzic2019} are important for hydrobaric bacteria and deep sea animals. In particular, the transition temperature changes by about +1$^\circ$\slash  40 bars in artificial lipids and bovine lung surfactant \cite{Ebel2001} and +1$^\circ$\slash  33 bars in \emph{E. coli} membranes \cite{Muzic2019}. 1--3$^\circ$\slash  100 bars was discussed in \cite{Fillafer2021}. The Mariana Trench is about 11 km deep and displays a hydrostatic pressure of about 1100 bars. Organisms have been found in the sediments at its bottom. 1100 bars would shift transitions by $\approx +28^\circ$.
\item voltage changes are relevant during the action potential and nerve excitation, and they can induce a fluid $\rightarrow$ gel transition at a value of 75 mV across the membrane (dependent on temperature) \cite{Thomas2022}.
\item changes in membrane composition such as the presence of anesthetics during surgery lower transition temperature \cite{Heimburg2007c, Graesboll2014}, and so does the activity of lipases that hydrolyze lipids into other lipid components.
\end{itemize}

Thus, there exist transient conditions where biological membranes are not fluid, and this gives rise to numerous phenomena, e.g.

\begin{itemize}
\setlength\itemsep{-0.1cm}
\item phase behavior
\item lipid and protein sorting by the host matrix
\item adaptation of the elastic constants
\item the possibility of pulse propagation in the biomembrane
\item the formation of membrane pores and sudden changes in permeability.
\end{itemize}

Since the changes in physical state close to the melting transitions are nonlinear, a relatively small perturbation may cause a large change in membrane state. We will call some of these phenomena ``excitability''. The title of this contribution ``The excitable fluid mosaic'' gives credit to the very valuable description of Singer \& Nicolson \cite{Singer1972} of the biomembrane in a steady state. However, we will add a layer of excitable phenomena to it that arise when the conditions in the membrane deviate from the steady state. The Singer \& Nicolson membrane is in the free energy minimum (the equilibrium state according to \cite{Singer1972}), and any perturbation (or excitation) will shift the state of the membrane away from equilibrium.

A further interesting phenomenon is adaptation of the membrane composition, which (as mentioned above) was already discussed by Singer \& Nicolson. The lipid composition of biological membranes can adapt to changes in the intensive variables. For instance, it is known that hydrobaric bacteria adapt the ratio between saturated and unsaturated lipids upon changes in pressure \cite{DeLong1985, Kaneshiro1995}. Similarly, trouts have a different ration of unsaturated to saturated lipids in winter and summer \cite{Hazel1979}. Further, the melting temperature of \emph{E.coli} bacteria adapts to changes in growth temperature \cite{Muzic2019}. The ratio between saturated and unsaturated lipids is important because unsaturated lipids generally melt at much lower temperature than saturated lipids. For instance, the lipid DOPC displays a melting temperature around -20$^\circ$C while DSPC displays a melting temperature of 55$^\circ$C. Both lipids have the same head group and the same chain length, but they differ in the number of double bonds in their chains. Adaptation was the main reason for Singer \& Nicolson to assume that under the physiologically relevant conditions the membrane is fluid and that melting of membranes can be dismissed as a factor that is important for membrane function.


\subsection{Phase diagrams}
\label{phasediagrams}

\begin{figure}[htbp]
\centering
\includegraphics[width=200pt]{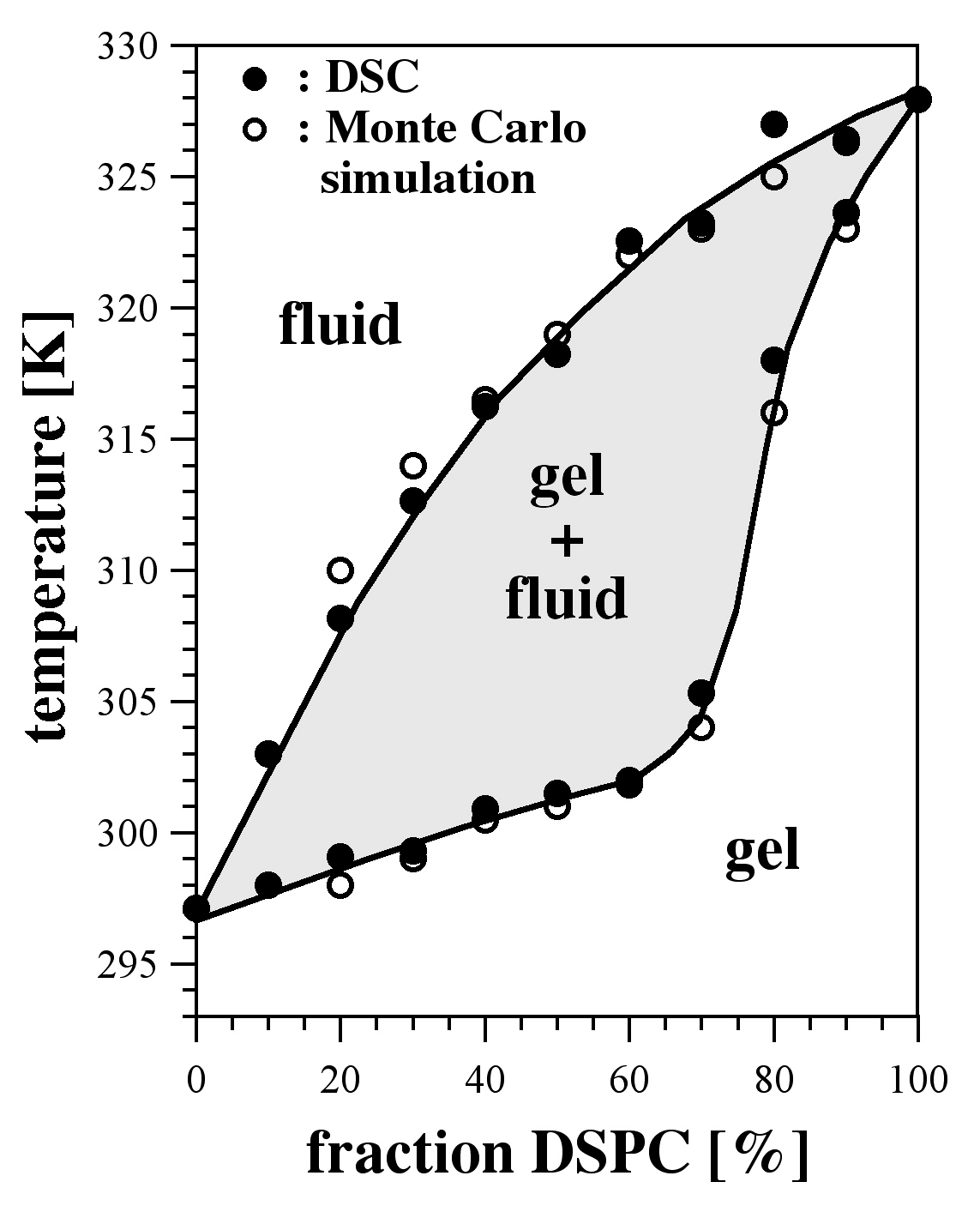}
\caption{Phase diagram of a mixture of DMPC and DSPC with individual melting temperatures of 297 K (23.8$^\circ$C) and 328 K (54.8$^\circ$C), respectively. The solid lines are the fluidus and solidus lines, the grey-shaded area corresponds to the regime where gel and fluid regions coexist. Closed symbols are from calorimetric profiles, while open symbols originate from Monte-Carlo simulations. From \cite{Hac2005}.}
\label{dmpc_dspc_a1}
\end{figure}

The biological membrane consists of a complex mixture of hundreds of different lipids and integral proteins. The lipids display differences in chain saturation, head group and the length of their fatty acid chains. Due to the difference in their chemical structure, they also display different melting temperatures. Mixtures of such lipids form regions within the membrane plane called phases or domains that display different local composition and fluidity. The simple example of a phase diagram of a binary lipid mixture is shown in Fig. \ref{dmpc_dspc_a1} (from \cite{Hac2005}) that shows temperature versus the composition of the membrane. One can see that there exists regions in the phase diagram that are purely fluid or purely gel, but other regions where one finds a separation of coexisting gel and fluid phases with different composition. The thermodynamics of phase behavior in membranes is thoroughly derived in \cite{Lee1977}. It is shown that one may not only find coexistence between gel and fluid regions, but also the coexistence of two fluid regions of different composition, or two gel regions of different composition. Thus, the fact that a membrane is fluid does not necessarily indicate that it is homogeneous. It can contains fluid segment of different composition.

\subsection{Domains}
\label{domains}

\begin{figure}[htbp]
\centering
\includegraphics[width=250pt]{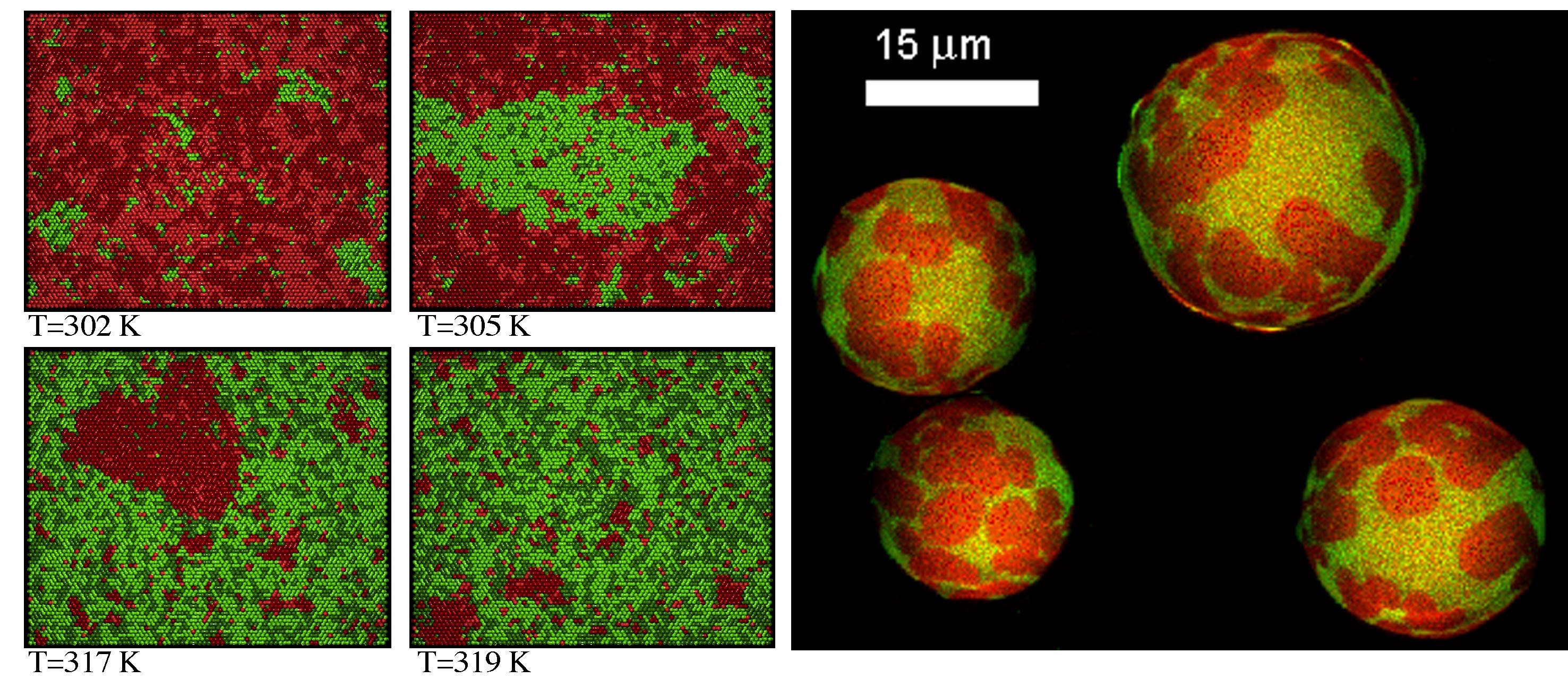}
\caption{Domain formation in lipid mixtures. Left: Monte Carlo simulations of a DMPC:DSPC=50:50 mixture. Red corresponds to a gel state and green to a fluid state. Dark shades of each color correspond to DSPC and bright shades to DMPC. Right: Fluorescence microscopy images from a DLPC:DPPC=30:70 mixture. From \cite{Hac2005}.}
\label{dmpc_dspc_b}
\end{figure}

A consequence of phase behavior is the formation of domains. Fig. \ref{dmpc_dspc_a1} shows the phase diagram of a DMPC-DSPC mixture as determined from calorimetric measurements (solid circles) and Monte-Carlo simulations (open circles) \cite{Hac2005}. Monte-Carlo simulations have the advantage over the usual construction of phase diagrams that it allows to include the interfacial energy between different lipids (first described in \cite{Doniach1978, Sperotto1991, Sugar1994, Heimburg1996a}). Including interfacial free energies leads to the possibility to find finite-size domains. Fig. \ref{dmpc_dspc_b} (left) shows simulated Monte-Carlo snapshots of domain formation at different temperatures in a membrane made of the DMPC-DSPC lipid mixture shown in Fig. \ref{dmpc_dspc_a1}. The difference between phase and domains is that phases are macroscopically separated and the interfacial energy between the phases can be neglected, which is not the case for domain formation where the interface cannot be neglected. Domains alter their size and composition as a function of temperature and composition. Gibb's phase rule discussed in section \ref{rafts} cannot be applied. Thus, domain formation and phase separation is not exactly the same thing. While phases have a well-defined composition, domains of different size also display different composition due to the free energy contribution originating from the domain interfaces. Fig. \ref{dmpc_dspc_b} (right) shows confocal fluorescence microscopy images of a DLPC-DPPC mixture that illustrates domain formation in giant vesicles \cite{Hac2005}. Generally, in the melting regime one finds coexistence of gel and fluid domains.

\subsection{The mattress model}
\label{themattressmodel}

\begin{figure}[htbp]
\centering
\includegraphics[width=200pt]{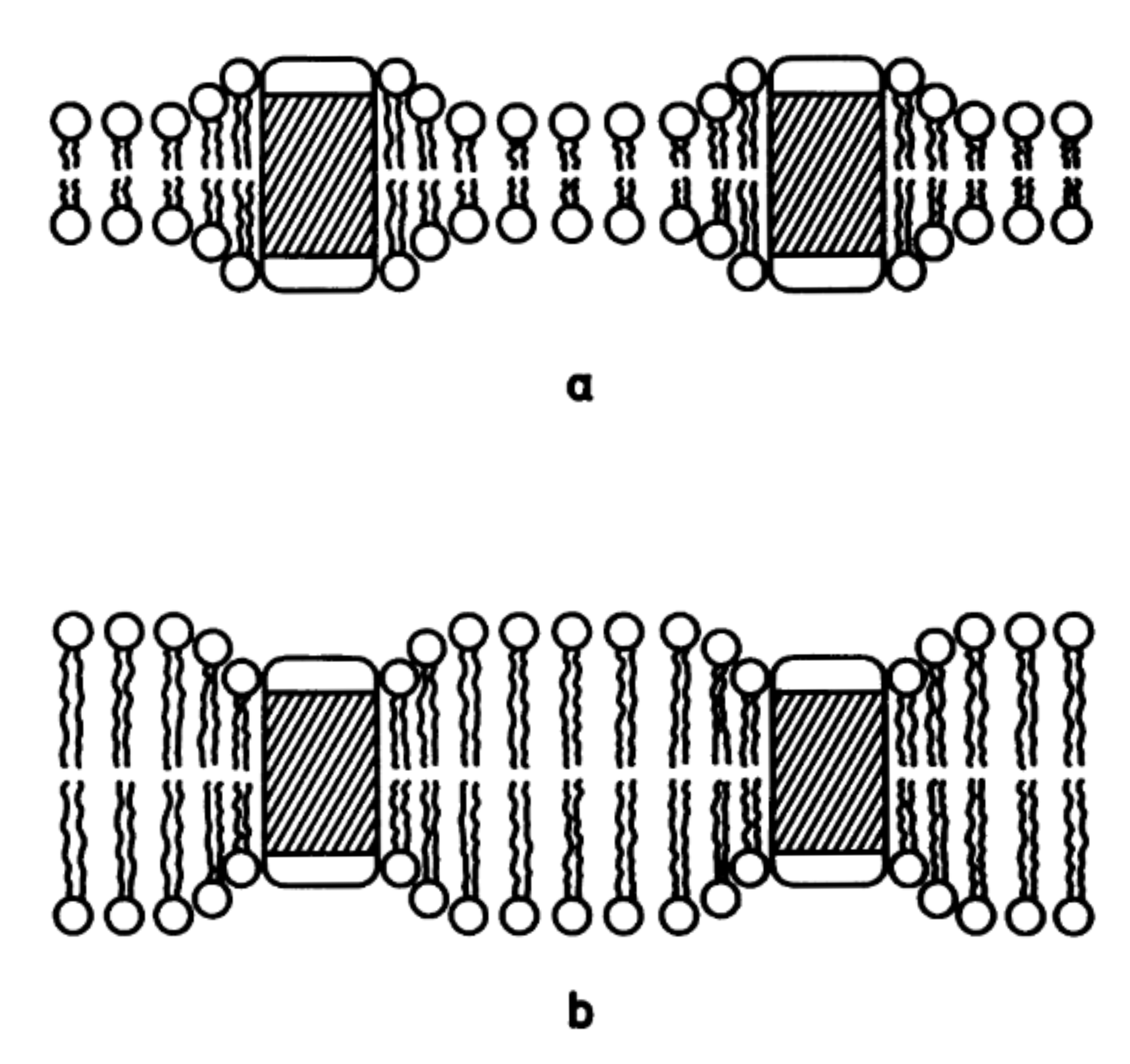}
\caption{The mattress model by Mouritsen \& Bloom suggests lipid-protein interaction that mediate forces between proteins. In the above schematic cartoon of integral proteins in the membrane one finds attractive forces between the proteins in both cases. From \cite{Mouritsen1984}.}
\label{mouritsenbloom}
\end{figure}

In 1984, Mouritsen and Bloom proposed the mattress model for integral proteins \cite{Mouritsen1984}. It discusses the interaction of integral proteins with gel and fluid lipid domains. It suggests that the thickness of the hydrophobic core of integral proteins and the membrane may be different. A gel membrane made of the lipid dipalitoyl phosphatidylcholine is thicker than a fluid membrane by about 17\% in \cite{Heimburg1998}. Thus, an integral protein that fits well into a fluid phase cannot fit into a gel phase membrane. As a consequence, the lipids at the interface of the proteins may have to adapt their length to the protein in order to minimize hydrophobic interactions as shown in Fig. \ref{mouritsenbloom}. The deformation of the lipids creates capillary forces that can lead to aggregation of proteins. This implies that in contrast to the assumptions of Singer \& Nicolson, the lipids of the membrane play an active role in the arrangement of proteins, and changes in both lipid state or in the length of the hydrophobic core of the integral protein will lead to rearrangements of the proteins. Since integral proteins must interact differently with gel and fluid membranes, they must influence the melting behavior of membranes which was also dismissed as a likely possibility in Singer \& Nicolson 's paper.

The influence of both peripheral and integral proteins on phase behavior and protein sorting was discussed in \cite{Heimburg1996a}. This paper confirms the validity of the mattress model and further suggests that peripheral proteins may bind differently to gel and fluid domains and cause changes in phase behavior, a possibility not considered by Singer \& Nicolson.


\subsection{Active proteins}
\label{activeproteins}

\begin{figure}[htbp]
\centering
\includegraphics[width=225pt,height=37pt]{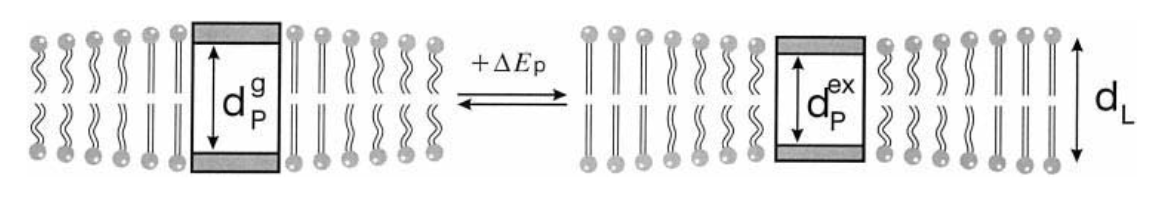}
\caption{Change in the hydrophobic length of an integral protein in a lipid membrane after an excitation process that changes the hydrophobic length of the protein. From \cite{Sabra1998a}.}
\label{sabra1998a}
\end{figure}

An interesting possible application of the mattress model proposed by Sabra \& Mouritsen are active ingredients, which could be an integral protein that changes its conformation after an activation process \cite{Sabra1998a, Sabra1998b}, see Fig. \ref{sabra1998a}. The activated protein prefers a different lipid environment and therefore both lipids and proteins will distribute differently before and after activation.

\begin{figure}[htbp]
\centering
\includegraphics[width=250pt]{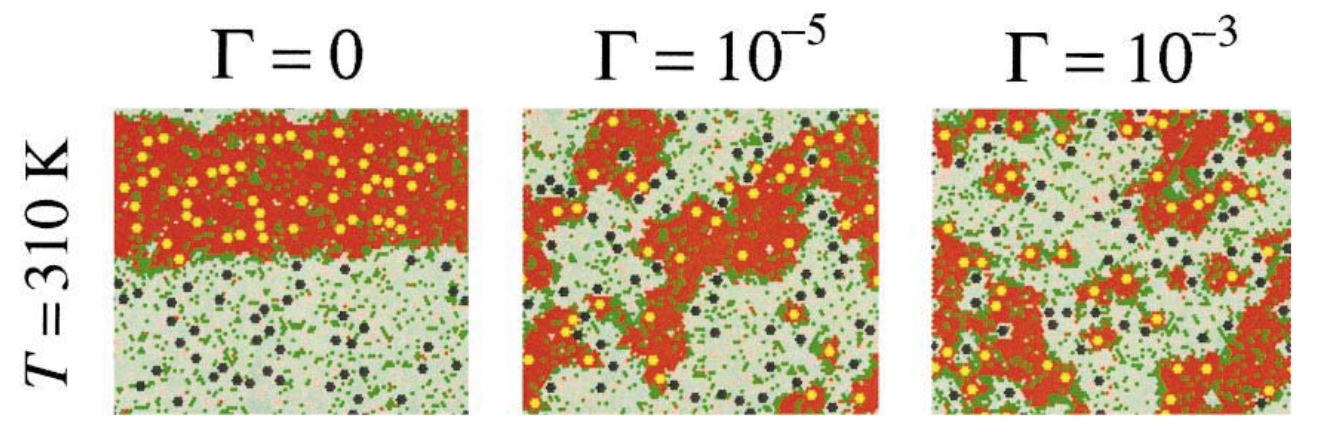}
\caption{Change in the arrangement of domains and proteins for different frequencies of protein activation processes. The result of the activations is a redistribution of both proteins and lipids. From \cite{Sabra1998a}.}
\label{sabra1998b}
\end{figure}

Fig. \ref{sabra1998b} shows a Monte-Carlo simulation of a DMPC-DSPC mixture containing active ingredients as shown in Fig. \ref{sabra1998a} \cite{Sabra1998a}. Randomly chosen proteins are activated by moving them from an elongated state into a state with a short hydrophobic core. With some time constant, the protein relaxes back into its original state. The mean arrangement of lipids and proteins now depends on how often the protein is activated. In Fig. \ref{sabra1998b} the arrangement of lipids and proteins is shown for different activation rates. It is obvious that the lipid-mediated interactions have an influence of the matrix.

\begin{figure}[htbp]
\centering
\includegraphics[width=200pt]{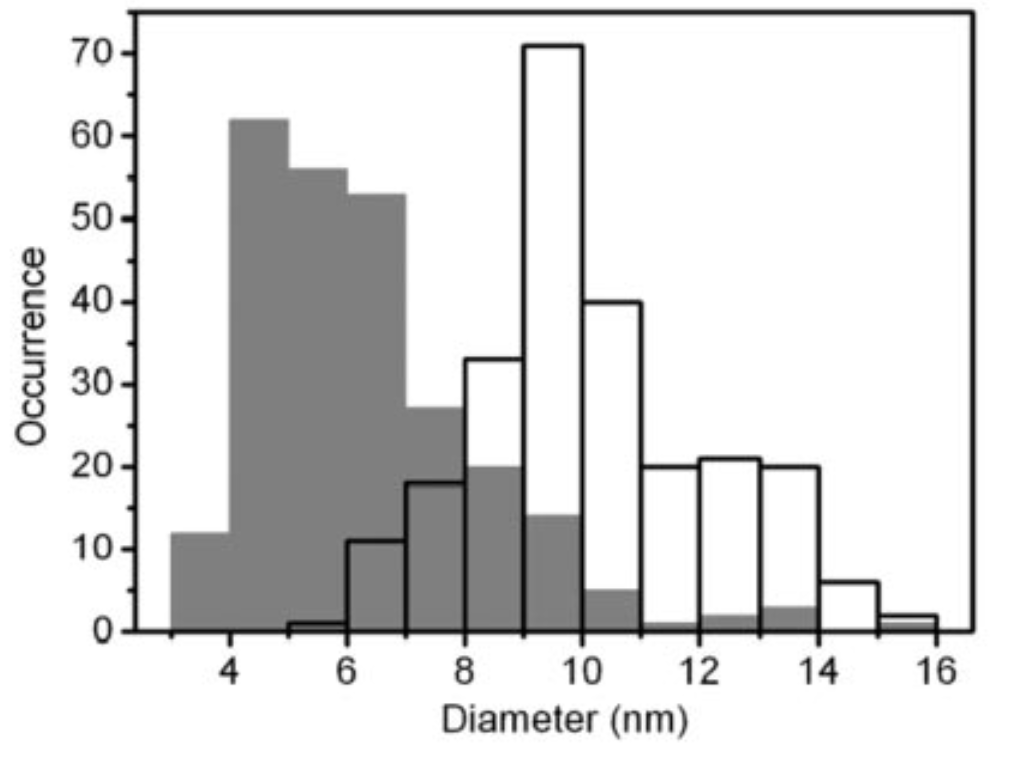}
\caption{Distribution of bacteriorhodopsin clusters in DPPC membranes in the absence of light (dark bars) and under light exposure (white bars). From \cite{Kahya2002}.}
\label{kahya2002}
\end{figure}

While the papers of Sabra and Mouritsen \cite{Sabra1998a, Sabra1998b} are of theoretical nature, there may exist proteins that in fact make use of such a concept. Sabra and Mouritsen pointed out that bacteriorhodopsin may be an example for such behavior and pointed at various demonstrations in the literature (e.g., \cite{Dumas1997, Dumas1999}). Kahya et al. \cite{Kahya2002} described the clustering of bacteriorhodopsin upon exposure to light by freeze-fracture electron microscopy. Fig. \ref{kahya2002} shows that the distribution of bacteriorhodopsin clusters in DPPC model membranes is different in the light-activated as compared to the dark state. Thus, the activation by light changes the distribution of the proteins, which is mediated by the membrane lipids.

\subsection{Rafts}
\label{rafts}

Biological membranes consist of hundreds of components (both lipids and proteins) and consequentially they display a rich phase behavior. In order to obtain a feeling for how many phases can coexist in a mixture, it is useful to consider Gibb's phase rule that states for a system at constant pressure that
\begin{equation}
F=K-J+1 \;,
\label{eq:rafts1}
\end{equation}
where $K$ is the number of chemical components, $J$ is the number of coexisting phases, and $F$ is the number of degrees of freedom. $F$ describes how many variables can be changed independently without changing the number of coexisting phases. For instance, in the diagram in Fig. \ref{dmpc_dspc_a1} for two lipid components ($K=2$) there exist three regimes: A gel phase in which both composition and temperature can be freely changed ($F=2$) and it still remains a gel phase, and a fluid phase where the same is true. In the two-phase region, only one variable, e.g. temperature, is free. For each temperature, the composition of the phases is fixed and cannot be changed freely ($F=1$). In a diagram with two components, there may also exist a single point at which $F=0$ and three phases coexist. This is the maximum number of coexisting phases in a two-component system. In a biological membrane composed of 200 components (just as an arbitrary example), the maximum number of coexisting phases is 201 ($K=200$, $J=201$ and $F=0$). In a finite system, macroscopic phase separation is unlikely and one rather expects domains of different state and composition that exceed the number of possible phases because the length of the domain interfaces represent further degrees of freedom. For this reason, it is clear that a biological membrane is unlikely to be homogenous, especially when one is close to a melting transition. It is more appropriate to consider it as a displaying a fluctuating distribution of regions with different states and compositions.

\begin{figure}[htbp]
\centering
\includegraphics[width=225pt]{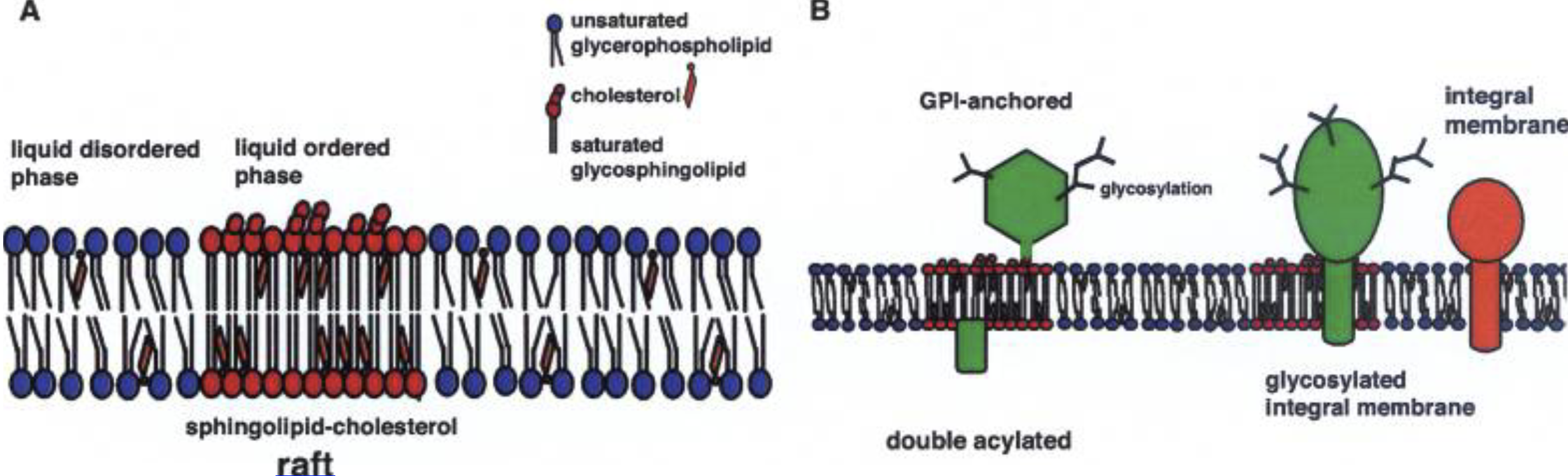}
\caption{Schematic drawing of lipid rafts in a biological membrane. The raft consists of regions enriched in saturated sphingolipids (with high melting temperature), cholesterol and GPI-anchored proteins. From \cite{Bagnat2002}.}
\label{bagnatsimons2002}
\end{figure}

\begin{figure}[htbp]
\centering
\includegraphics[width=225pt,height=109pt]{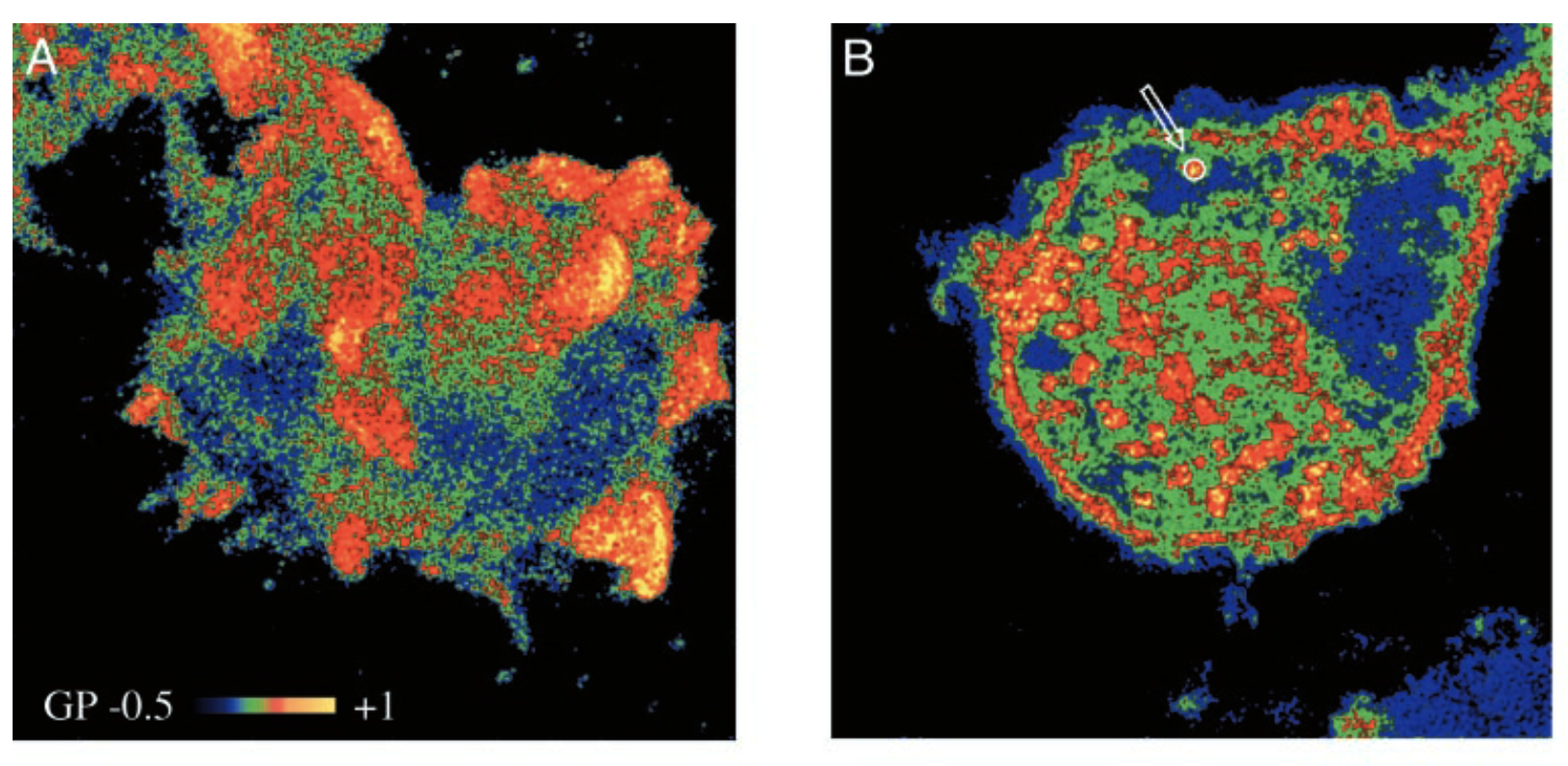}
\caption{Fluorescence microscopy images of domain formation in living cells. Left: A mouse macrophage (RAW264.7 cell, horizontal width 28.3 \textmu m) and Right: A human leukemia monocytes cell (THP-1, horizontal width 39 \textmu m). Different colors indicate general polarization values indicative for different lipid states. Yellow\slash red shades indicate more solid regions while blue\slash green shades indicate more liquid regions. From \cite{Gaus2003}.}
\label{gaus2003}
\end{figure}

In fact, it has been proposed that the biological membrane contains lipid-protein platforms called rafts \cite{Simons1997}. Meanwhile there exists a large literature on raft formation that agrees in that they are enriched in sphingolipids with high melting temperature, cholesterol and certain lipid-anchored proteins. A schematic drawing is given in Fig. \ref{bagnatsimons2002} (from \cite{Bagnat2002}). Such domains are also found in fluorescence microscopy experiments. An example is given in Fig. \ref{gaus2003} for macrophage and leukemia monocyte cells. The different colors in the images reflect the order of the lipid membrane. One can see gel domains in red\slash yellow shades and fluid domains in blue\slash green shades. Nice domain structures in living cells were also reported by \cite{Harder1998}. While in the early phase of raft research they were rather considered as molecular platforms with a particular function (like ribosome that are complexes of proteins and RNA with a particular function), rafts are these days mostly seen as gel-like domains that form as a consequence of phase behavior of the membranes. Thus, the composition of such domains most likely largely depends on composition and temperature, but probably also of other variables such as local pH and voltage gradients.

\subsection{Bending}
\label{bending}

The microscopic image in Fig. \ref{gaus2003} (left) also show another possible feature of membranes: curvature. In 1973, Helfrich published a famous paper that treats the membrane as a two-dimensional liquid crystal and assigns elastic constants to its deformation free energy \cite{Helfrich1973}. For a two-dimensional fluid membrane in which the elastic constants are uniform and independent of orientation, the elastic free energy density is given by
\begin{equation}
g=\frac{1}{2}K_B\left(c_1+c_2-c_0\right)^2+K_G \;c_1c_2. \;,
\label{eq:bending1}
\end{equation}
where $c_1$ and $c_2$ are principle curvatures, and $c_0$ is the spontaneous curvature which is the equilibrium curvature of the membrane when no forces are present. $K_B$ is the bending modulus and $K_G$ is the Gaussian modulus. The total free energy of curvature of a closed surface is $G=\oint g dA$. One can add other terms to eq. (\ref{eq:bending1}), for instance when the membrane is stretched or sheared.

\begin{figure}[htbp]
\centering
\includegraphics[width=250pt]{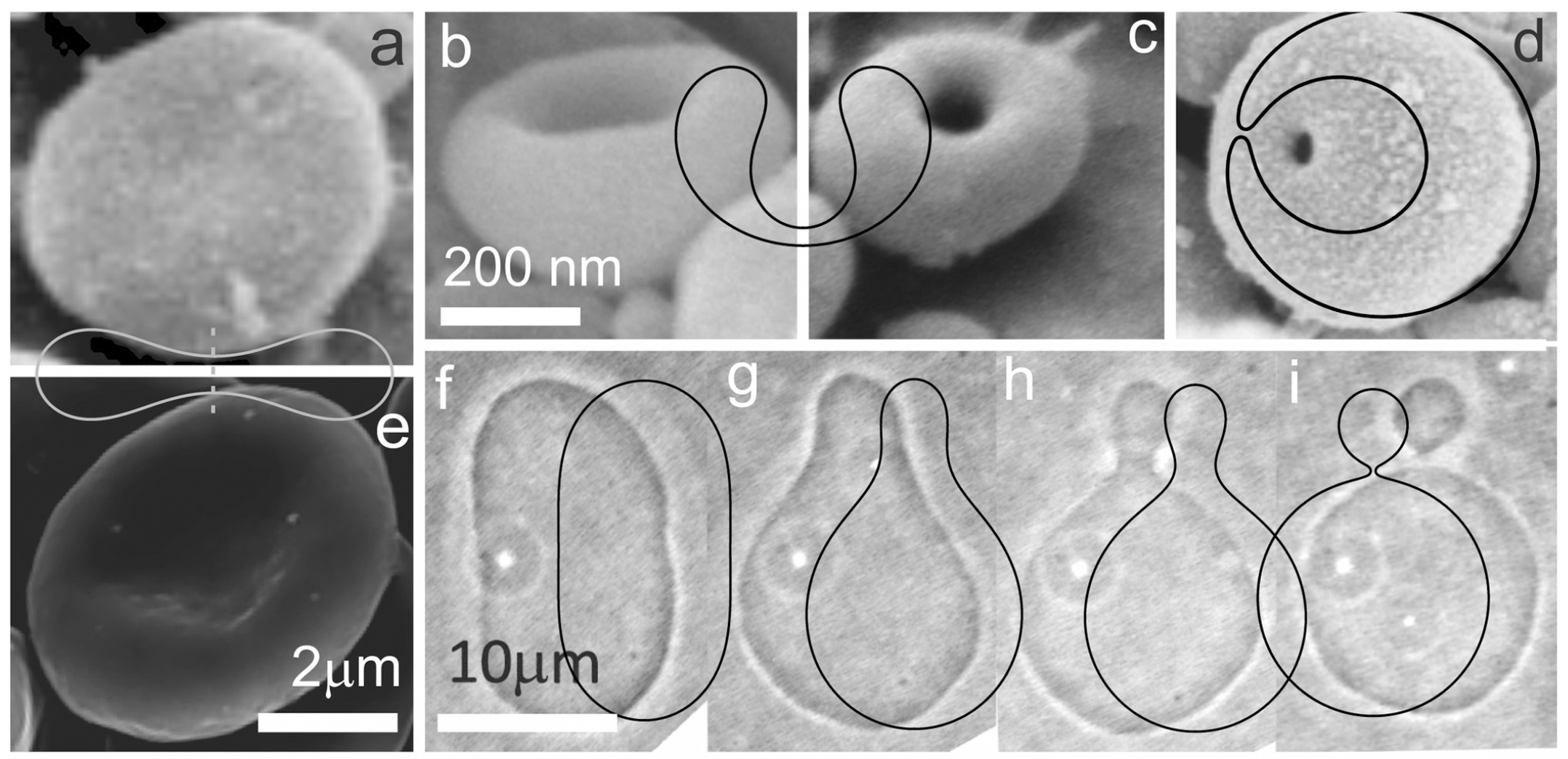}
\caption{Cell membrane nano-vesicles imaged by electron microscopy (a-e) and lipid vesicles imaged by light microscopy (f-i) and shapes minimizing the free energy (solid lines). From \cite{KraljIglic2020}. }
\label{iglic2020}
\end{figure}

Examples for the application of membrane curvature are vesicular shapes \cite{Berndl1990, Lipowsky1995, KraljIglic2020}, adhesion phenomena \cite{Sackmann2002, Weikl2002} and the calculation of the energy of fusion pores \cite{Kozlovsky2002}. Fig. \ref{iglic2020} shows several cell membrane nano vesicles (panels a-e) and lipid vesicles (f-i), and the corresponding minimum free energy shapes (from \cite{KraljIglic2020}). Interestingly, the different vesicular shapes in i-f had been generated by slightly adjusting temperature (and thus membrane area), see also \cite{Seifert1997}.

Elastic behavior was not considered in the fluid mosaic model. However, the elastic behavior of membranes is important for everything that is related to change of the shape of cells, vesicles and the membrane surface. If there were thermodynamic ways to control the magnitude of the elastic constant, one would provide a possibility to influence the shapes of biological membranes. If these changes occur locally, one would provide means to create local mechanical perturbations.


\subsection{Fluctuations and elastic constant}
\label{fluctuationsandelasticconstant}

\begin{figure}[htbp]
\centering
\includegraphics[width=250pt]{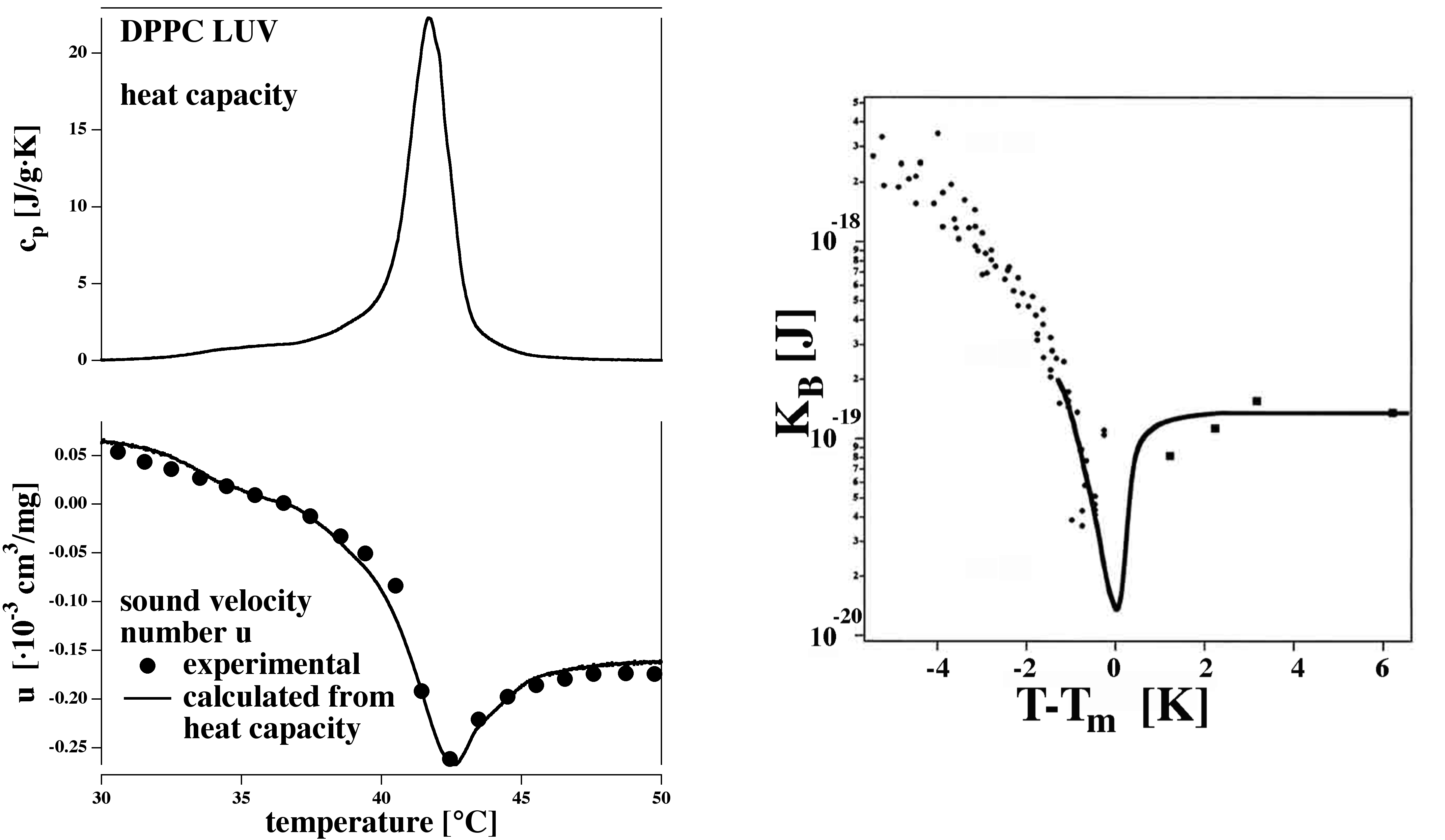}
\caption{Left: Sound velocity at 5 MHz for DPPC membranes. Top: Heat capacity profile. Bottom: Experimental sound velocity number (symbols) and a profile calculated from the heat capacity (solid line). From \cite{Halstenberg1998, Heimburg2007a}. Right: Change of the bending elasticity of DPPC membranes calculated from fluctuation theory and measured by \cite{Dimova2000, Meleard1997}. Figure from \cite{Heimburg2007a}.}
\label{elastic}
\end{figure}

So far we have mostly discussed the distribution of molecules in phases and domains. However, there exists another interesting layer on top of the arrangement of molecules that does not receive the attention that it deserves: Thermal fluctuations. It is clear from what has been stated above that domains may have variable size, state and composition. There exists a distribution of the extensive variables around an average state. The mean square deviation of an extensive variable from the average state reflects the magnitude of the fluctuations of the system. This becomes important for a property of membranes that has not been touched so far and has also not received any notice in Singer \& Nicolson's model: the susceptibilities. A susceptibility is a measure for how large the change of an extensive property of the membranes is when an intensive variable is changed. We have already mentioned the heat capacity at constant pressure, $c_p=(\partial H/\partial T)$, where the enthalpy $H$ is the extensive variable, and the temperature $T$ is the intensive variable. At the melting temperature, a small change in temperature causes a large change in enthalpy, i.e., the membrane enthalpy is very susceptible to small temperature changes. Another important susceptibility is the isothermal lateral compressibility $\kappa_T^A=\left<A\right>^{-1}(\partial A/\partial \Pi)_T$, where the membrane area $A$ is the extensive variable and the lateral pressure $\Pi$ is the intensive variable. At the melting temperature, the membrane area is very susceptible to small changes in lateral pressure, i.e., it is very compressible and the compression modulus is small. Interestingly, all susceptibilities are related to a fluctuation relation \cite{Heimburg1998}. For heat capacity and lateral compressibility these are
\begin{eqnarray}
c_p&=&\frac{\left<H^2\right>-\left<H\right>^2}{RT^2}\nonumber\\
	\kappa_T^A&=&\frac{\left<A^2\right>-\left<A\right>^2}{\left<A\right>RT}\;,
\label{eq:fluct1}
\end{eqnarray}
meaning that the mean square fluctuations of the enthalpy are proportional to the heat capacity and the mean square fluctuations in area are proportional to the area compressibility. Similar fluctuation relations exist for the capacitance, which is proportional to fluctuations in charge \cite{Heimburg2012}, and bending elasticity, which is related to the fluctuations in curvature \cite{Zeman1990, Strey1995}. The importance of such fluctuation-relations cannot be overestimated. If fluctuations are high, the conjugated susceptibility is also high. In the heat capacity maximum of a melting transition, the enthalpy fluctuations are high and so are the area fluctuations. Generally, membranes in transitions are softer in the transition range as compared to both gel and fluid phase, and can be stretched and bent easily. Practically, it has been shown that the elastic constants are closely related to the heat capacity of the membrane \cite{Heimburg1998, Ebel2001, Heimburg2007a, Muzic2019} and that it is possible to calculate elastic constants from the heat capacity. Since the heat capacity is a function of temperature, so are the elastic constants. In other words: Elastic constants are not truly constant but depend on the state of the membrane and the intensive variables. Two examples are given in Fig. \ref{elastic}.

It has been shown experimentally that the area compressibility of membranes is high in the melting transition \cite{Evans1982}, and so is the volume compressibility \cite{Halstenberg1998, Ebel2001, Schrader2001}. Further, the bending elasticity is also high, and the bending modulus $K_B$ is small \cite{Meleard1997, Dimova2000}. Fig. \ref{elastic} (left, top) shows the heat capacity of DPPC vesicles, and the sound velocity of dispersions of such lipids. The sound velocity $c$ is related the compressibility via $c=\sqrt{1/\kappa_S^V\cdot \rho^V}$, where $\kappa_S^V$ is the adiabatic volume compressibility, and $\rho^V$ is the mass density. Fig. \ref{elastic} (left, bottom) shows the sound velocity number $u=(c-c_{H_2O})/c_{H_2O}[L]$, which is a measure for the change in sound velocity normalized by the speed of sound in water $c_{H_2O}$ and the lipid concentrations {[L]}. It can be seen that in the lipid transition, the sound velocity of a vesicular dispersion is at a minimum. The solid line represents the prediction from the heat capacity profile shown in Fig. \ref{elastic} (left, top). Fig. \ref{elastic} (right) shows the bending modulus $K_B=1/\kappa_B$ as determined in experiments (symbols) and calculated from the heat capacity (solid line). The predictions describe the experimental profiles well and thus there exists a proof of concept for the relation between heat capacity and the elastic constants, and the respective fluctuations of the conjugated extensive variable.

Extrapolating to biological membranes, the result of this section is that they are softer at temperatures below body temperature. Any change inside of a cell, e.g., a sudden change in pH, pressure or temperature, will change the elastic constants of membranes. This generates many possibilities for the controll of membrane function. Below we will discuss two of them: the generation of sound pulses in membrane cylinders which are a consequence of the larger compressibility of the membranes below body temperature, and the control of membrane permeability and the emergence of channel events in artificial and biological membranes.


\subsection{Pulse propagation}
\label{pulsepropagation}

\begin{figure}[htbp]
\centering
\includegraphics[width=250pt]{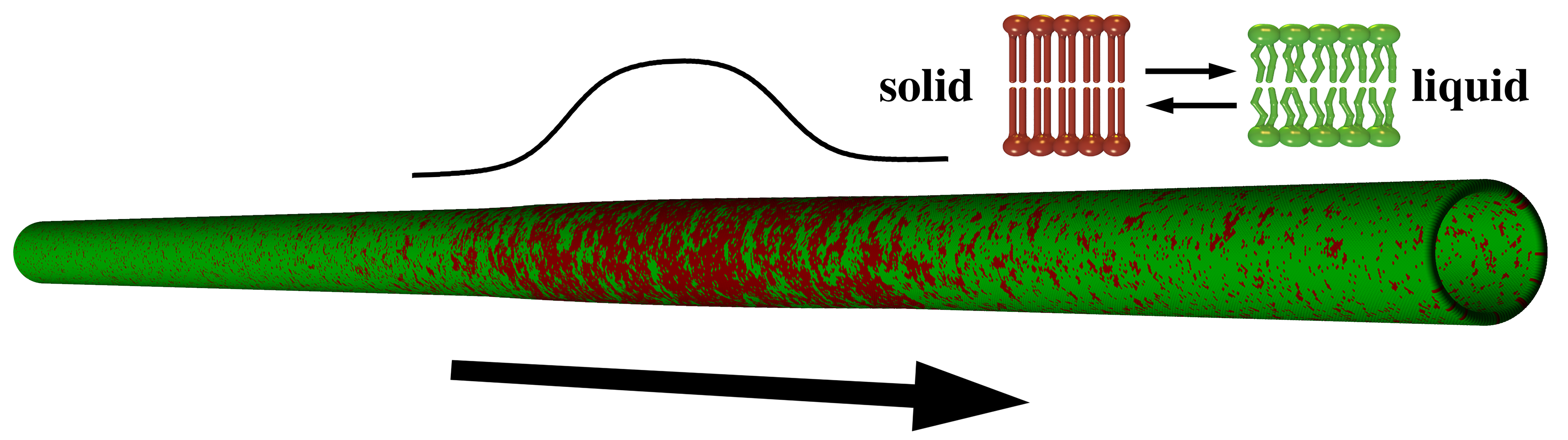}
\caption{Schematic image of a soliton propagating in a membrane cylinder made of fluid lipids. The pulse is rather gel-like. From \cite{GonzalezPerez2016}.}
\label{gonzalezperez2016}
\end{figure}

An application of elastic and fluctuation theory is the sound propagation within membrane cylinders or axons. The possibility of propagating electromechanical pulses in membranes was described by several authors, e.g., \cite{Heimburg2005c, Andersen2009, Lautrup2011, Griesbauer2012a, Griesbauer2012b, Shrivastava2014, Kappler2017, Schneider2021}.
As described above, the sound velocity is related to the compressibility of the membrane. This is also true for the lateral sound velocity in the membrane and the lateral compressibility of the membrane. The sound velocity displays a minimum at the melting transition (Fig. \ref{elastic}). If the membrane is in its fluid state slightly above a transition (as described by Singer \& Nicolson), one finds the possibility of the propagation of pulses called solitons. The wave equation in the presence of dispersion is given by \cite{Heimburg2005c, Andersen2009, Lautrup2011}
\begin{equation}
\frac{\partial^2 \rho^A}{\partial t^2}=\frac{\partial}{\partial x}\left(c^2\frac{\partial \rho^A}{\partial x}\right)-h\frac{\partial^4 \rho^A}{\partial x^4}\;,
\label{eq:pulse1}
\end{equation}
where $\rho^A$ is the area mass density in the membrane, $c$ is the 2 dimensional density dependent sound velocity in the membrane, and $h$ is a constant describing the magnitude of dispersion (the frequency-dependence of the sound velocity). The left half of this equation (without the dispersion term containing the constant $h$) is just the one-dimensional wave equation from hydrodynamics in the absence of friction. When entering the experimentally determined sound velocity of a membrane close to the melting transition into eq.(\ref{eq:pulse1}), it displays a (local) solitary solution which was called `a soliton' \cite{Heimburg2005c}. The soliton consists of a local gel-like (or solid) region (with larger area density) traveling in a fluid (or liquid) membrane with a velocity that is somewhat smaller than the sound velocity (about 1--100 m\slash s), see Fig. \ref{gonzalezperez2016}. In \cite{Lautrup2011} it was shown that the solitons are stable even in the presence of minor friction which is present in all viscous environments.

\begin{figure}[htbp]
\centering
\includegraphics[width=250pt]{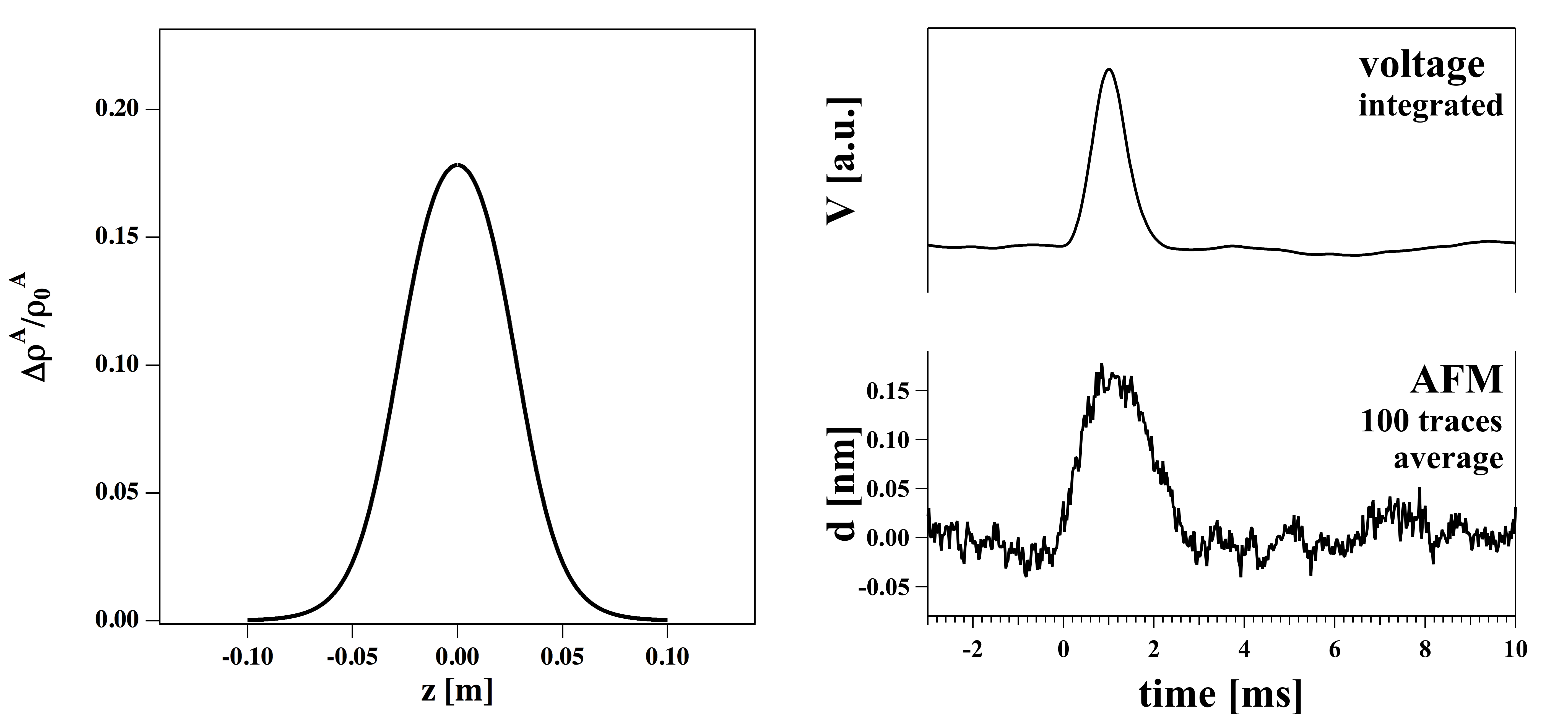}
\caption{Left: Soliton in a cylindrical axon along the coordinate $z$. Adapted from \cite{Heimburg2005c}. Atomic force microscopy measurement of a propagating thickness-pulse in a neuron from lobster connectives and the corresponding voltage trace. Adapted from \cite{GonzalezPerez2016}.}
\label{soliton}
\end{figure}

A calculated soliton profile is shown in Fig. \ref{soliton} (left). It displays a remarkable similarity with the nerve pulse. Its velocity is close to that reported for nerve pulses. One finds that the nerve membrane displays a thickness change both theoretically and in the experiment \cite{Iwasa1980a, Iwasa1980b, Tasaki1989, Kim2007, GonzalezPerez2016, Ling2020}. The voltage and thickness changes during the action potential in a lobster neuron are shown in Fig. \ref{soliton} (right). Furthermore, during a nerve pulse one finds a reversible heat exchange between the nerve and its aqueous environment \cite{Abbott1958, Howarth1968, Howarth1975, Ritchie1985, Tasaki1989}, which finds its explanation in the latent heat of the transition that is reversibly released and reabsorbed by the nerve membrane \cite{Heimburg2021}.

\subsection{Channels}
\label{channels}

\begin{figure}[htbp]
\centering
\includegraphics[width=200pt]{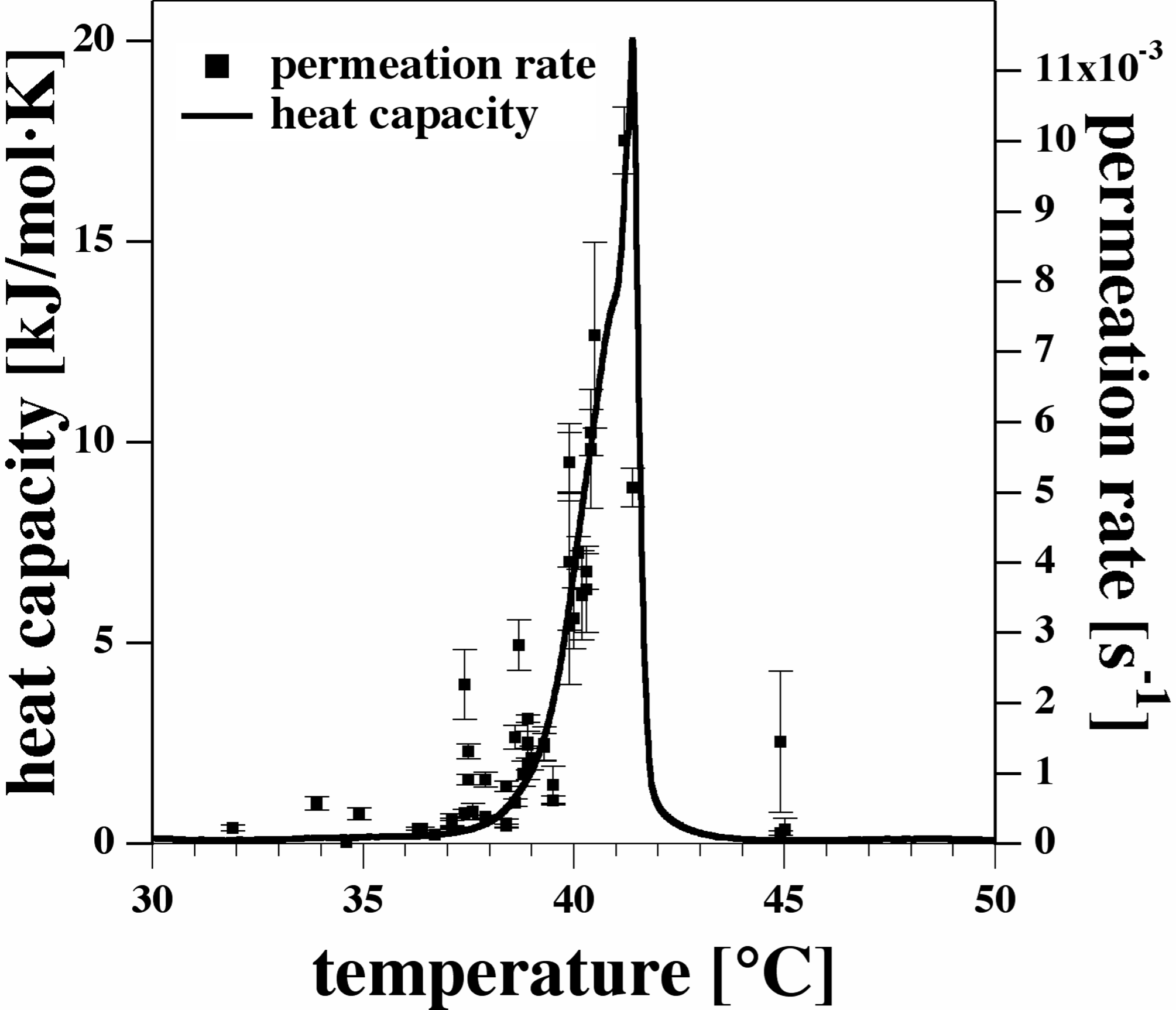}
\caption{Permeability of a DPPC:DPPG=95:5 membrane for a fluorescence dye as measured by fluorescence correlation spectroscopy (symbols). The permeability is compared to the heat capacity profile (solid line). From \cite{Blicher2009}.}
\label{blicher2009}
\end{figure}

It is well-known that membranes become permeable in the melting transition \cite{Papahadjopoulos1973, Sabra1996, Blicher2009}. This has been explained by the increase in the lateral compressibility in the transition regime. The free energy of pore formation is given by \cite{Nagle1978b}
\begin{equation}
\Delta F = \frac{1}{2 \kappa_T^A}\frac{\Delta A^2}{A_0}\;,
\label{eq:pore1}
\end{equation}
where $\Delta A$ is the area of the pore opening, and $A_0$ is the total area of the membrane. $\Delta F$ corresponds to the work that is necessary to create a pore. When that work is small ($\kappa_T^A$ is large) the likelihood to find an open pore is larger and the membrane is more permeable. For this reason, one finds a permeability maximum in the melting transition, see Fig. \ref{blicher2009}. The generation of such pores has been described by numerous authors, e.g., \cite{Glaser1988} and was also investigated with molecular dynamics simulations \cite{Boeckmann2008}.

\begin{figure}[htbp]
\centering
\includegraphics[width=250pt]{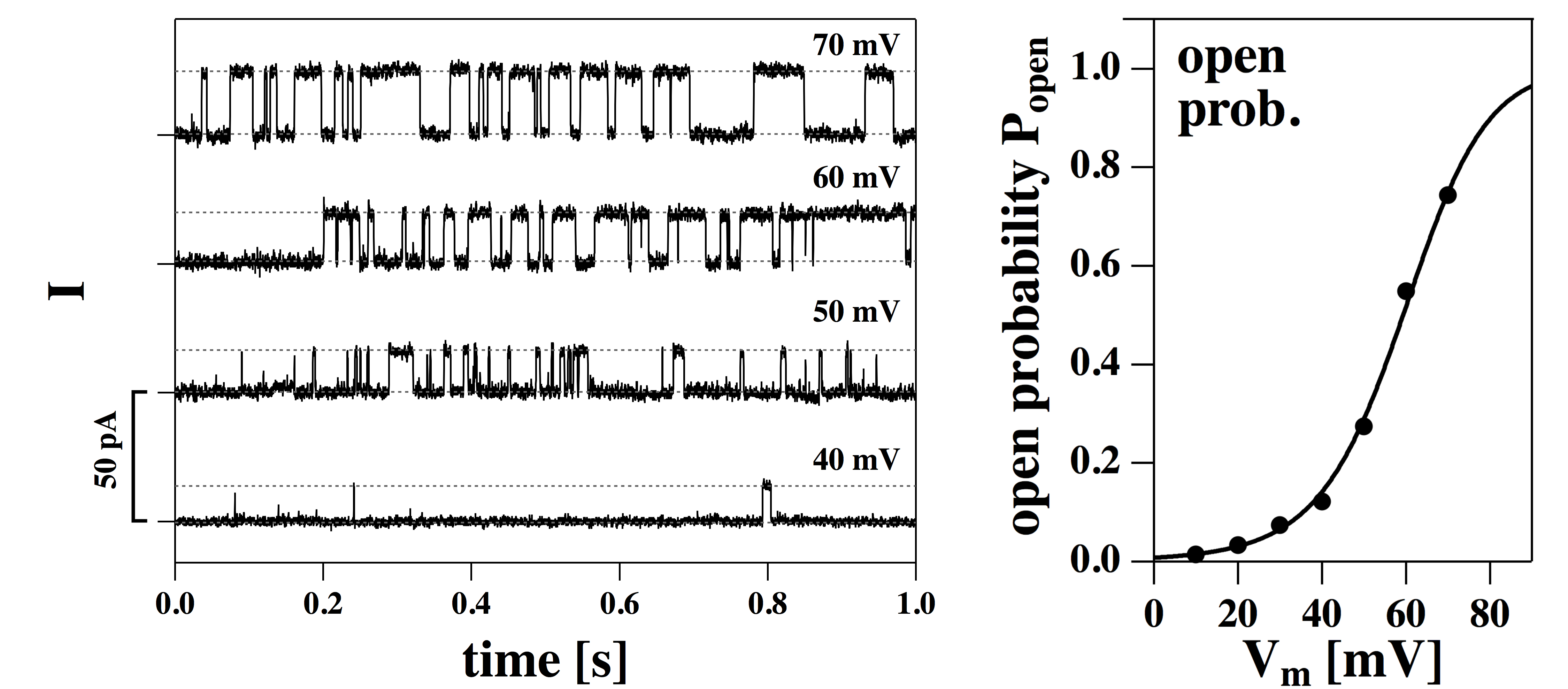}
\caption{Patch-clamp recordings of lipid ion channels in a model membrane consisting of DMPC:DLPC=10:1. The open probability displays voltage-gating with a threshold-voltage of around 50 mV. From \cite{Blicher2013}. }
\label{blicher2013}
\end{figure}

The permeability of membranes can also be measured in black lipid membrane experiments \cite{Antonov1980, Antonov2005, Blicher2009, Heimburg2010} and in patch clamp recordings \cite{Laub2012, Blicher2013, Zecchi2021}. Surprisingly, one finds single permeation events that strongly resemble ion channel recordings in biological membranes (Fig. \ref{blicher2013}). They share so many similarities with channel proteins that they were called `lipid ion channels' \cite{Heimburg2010}. In fact, from just analyzing the appearance of the channel traces, the channel events appear indistinguishable from protein channels \cite{Laub2012}. The probability to find an open lipid channel is both voltage-gated \cite{Laub2012, Blicher2013, Zecchi2021} and mechanosensitive \cite{Zecchi2021}, and naturally also temperature-sensitive and they can be `blocked' by anesthetics \cite{Blicher2009, Wodzinska2009}. Thus, the channel open-probability responds to changes in the intensive thermodynamic variables in a similar manner as reported to some protein channels.

Since there exists no single molecular objects in a pure lipid membrane that can create a pore or can be blocked, it is clear that close to transitions the membrane as a whole is a receptor for all of these variables (such as voltage, lateral tension or chemical potential of anesthetics). It is a thermodynamic property of the total system rather than being a property of a single molecule or a single pore. It is an open question how lipid channels and protein channels are related - or if they are in fact in some or all case identical or closely related (see discussion in \cite{Mosgaard2013b}). Clearly, the lipid channels are an emergent phenomenon close to transitions not considered by Singer \& Nicolson. In particular, these are phenomena that are lipid-mediated which makes it clear the lipids in a biological membranes are much more than an inert liquid. Instead, they are active players.


\section{Conclusions}
\label{conclusions}

The fluid mosaic model by Singer and Nicolson \cite{Singer1972} shows a remarkable insight into the nature of biological membranes at rest. It postulated the existence of integral proteins on thermodynamic grounds and proposed that the native membrane serves as a two-dimensional solvent in which proteins are free to diffuse. Thus, the ``fluid mosaic'' describes an ideal solution of proteins in a lipid solvent - and mass-action-like reactions between them. It discusses many modern aspects of membranes such as melting transitions, lipid-protein interactions and domain sizes. Singer \& Nicoslon came to the conclusion that the membranes are found in the fluid state above the transition. They dismiss lipid-protein interactions as small and do not attribute a role to the melting behavior of membranes and to membrane elasticity. We agree with many of these findings in a steady state situation where the membrane is not perturbed. However, when deviating from this state, biomembranes show a fascinating wealth of emergent properties. There exist several layers of phenomena on top of Singer \& Nicolson. These include

\begin{itemize}
\setlength\itemsep{-0.1cm}
\item phase behavior
\item domain and raft formation and sorting of proteins
\item membrane elasticity
\item fluctuations
\item electromechanical pulses resembling the nerve pulse
\item lipid channels, pores resembling ion channel proteins, and membrane permeability
\end{itemize}

Most or all of the evidence for their existence and relevance was published after the year of publication of the fluid mosaic model in 1972. The Singer \& Nicolson model rests on the assumption that something like a physiological condition exists, e.g., physiological temperature, physiological membrane tension, pH, electrical potential and ion concentrations. These conditions are such that the membrane is found in its fluid state where the fluid mosaic model probably is mostly an accurate description. However, we believe that constant physiological conditions do not exist. Deviations from steady state conditions via changes of the intensive variables (e.g., changes of the chemical potential of protons due to the action of hydrolytic enzymes, or temperature changes in cold-blooded animals), both locally and globally, may move the biomembrane into a physical regime where all these additional physical layers become important. Deviation from steady-state conditions can generate or inhibit a signal or an excitation of the membrane by either moving the conditions towards or away from a situation where transitions in the membrane become important (reviewed in \cite{Heimburg2022b}). This may lead to a thermodynamic control of the lateral arrangement of lipids and proteins in the membrane plane into domains and allow for a control of this arrangement via active proteins that change their shape upon excitation. It allows for the elastic constants to change vesicular shapes or generate exo - and endocytotic events. It further allows for the generation of electromechanical pulses or solitons in membrane that display a remarkable similarity with action potentials. Finally, these excited membrane states display fluctuations that lead to pore formation in the membrane with a remarkable similarity to protein ion channels. Thus, deviation from steady-state conditions open the door to a fascinating realm of excitation phenomena, for which the fluid mosaic model is just the starting point.

\vspace{0.3cm} \noindent
\textbf{Author contributions:}
TH designed and wrote the article.

\vspace{0.3cm} \noindent\textbf{Acknowledgments:} The fields of membrane phase behavior, domain formation and the elastic behavior of membranes (including the possibility of pulse propagation and channel formation) are huge. Our choice of references to the work in the field is far from being complete, and omissions of important contributions are by no means judgments about their importance.


\small{

}

\end{document}